%% file: Sentiment-SATD-V2.tex
\RequirePackage{fix-cm}
\documentclass[smallextended]{svjour3}

\makeatletter
\def\cl@chapter{\@elt {theorem}}
\makeatother

\input{0-packages}
\input{0-commands}
\smartqed  % flush right qed marks, e.g. at end of proof
\begin{document}

\title{Negativity in Self-Admitted Technical Debt: How Sentiment Influences Prioritization}



\author{Nathan Cassee         \and
        Neil Ernst \and
        Nicole Novielli \and
        Alexander Serebrenik %etc.
}


\institute{Nathan Cassee \at
                Eindhoven University of Technology, The Netherlands \\
                  \email{n.w.cassee@tue.nl}   
           \and
           Neil Ernst \at 
            University of Victoria, Canada\\
            \email{nernst@uvic.ca}  
           \and
               Nicole Novielli \at
               University of Bari, Italy\\
        \email{nicole.novielli@uniba.it}           %  \\
               \and
               Alexander Serebrenik \at
                Eindhoven University of Technology, The Netherlands\\
                  \email{a.serebrenik@tue.nl} 
}

\date{Received: date / Accepted: date}

\maketitle

\input{2-0-abstract}

\input{2-1-introduction}

\input{2-2-methodology}

\input{2-3-results}

\input{2-4-discussion}

\input{2-5-threats}

\input{2-6-related-work}

\input{2-7-conclusion}

%\begin{acknowledgements}
%\end{acknowledgements}

 \section*{Conflict of interest}

 The authors declare that they have no conflict of interest.

 \section*{Data Availability Statement}

 The data used for this study and the notebook used to analyze the data are publicly available in a figshare repository.\footnote{\url{https://doi.org/10.6084/m9.figshare.26863435}}

\bibliographystyle{spbasic}   

\bibliography{bibliography}

\end{document}

%% file: 0-packages.tex
\usepackage{amsmath,amsfonts}
\usepackage{algorithmic}
\usepackage{textcomp}
\usepackage{xcolor}
\usepackage{multirow}
\usepackage{xspace}
\usepackage{balance}

\usepackage{lmodern}
\usepackage{usebib}

\usepackage{enumerate}
\usepackage{verbatim}
\usepackage{booktabs}
\usepackage{multirow}

\usepackage{subcaption}

\usepackage{tabularray}
\usepackage{graphicx}

\usepackage{fontawesome}
\usepackage{amssymb}

\usepackage{cite,natbib}
\usepackage{rotating}
\usepackage[capitalise]{cleveref}
\usepackage{listings}

\usepackage{lstautogobble}
\usepackage{longtable}
\usepackage{tikz}

\usepackage{csquotes}
\usepackage{customtables}
\usepackage{subcaption}

\usepackage{tcolorbox}

\usepackage{pifont}
\usetikzlibrary{arrows}

%% file: 0-commands.tex
\newboolean{showcomments}
    
\setboolean{showcomments}{false}

\ifthenelse{\boolean{showcomments}}
  {\newcommand{\nb}[2]{
    \fbox{\bfseries\sffamily\scriptsize#1}
    {\sf\small$\blacktriangleright$\textit{#2}$\blacktriangleleft$}
   }
  }
  {\newcommand{\nb}[2]{}
  }

\newcommand{\todo}[1]{\textcolor{orange}{\nb{TODO}{\textbf{#1}}}\xspace}
\newcommand{\nc}[1]{\textcolor{magenta}{\nb{Nathan}{#1}}}
\newcommand{\as}[1]{\textcolor{teal}{\nb{Alexander}{#1}}}
\newcommand{\neil}[1]{\textcolor{blue}{\nb{Neil}{#1}}}

\newcommand{\etal}{{\emph{et al.}}\xspace}
\newcommand{\eg}{{\emph{e.g.},}\xspace}
\newcommand{\ie}{{\emph{i.e.},}\xspace}

\newcounter{rqcounter}

\newcommand{\researchq}[2]{
    \refstepcounter{rqcounter}\label{rq:#1}
    \begin{description}
        \item[\textit{RQ\textsubscript{\therqcounter}:}] \textit{#2} 
    \end{description}
}

\newcommand{\refrq}[1]{\emph{RQ\textsubscript{\ref{#1}}}\xspace}

\bibinput{bibliography}

\renewcommand{\cite}[1]{\citep{#1}}

\newcommand{\cmark}{\ding{51}}%
\newcommand{\xmark}{\ding{53}}%

\newcommand{\draft}[2]{
\ifnum #1>5
  \textcolor{blue}{#2}
\else 
#2
\fi
}

\definecolor{blueish}{cmyk}{61, 47, 0, 40}

\newcommand{\implication}[2]
{
\begin{tcolorbox}[colback=blueish!5!white,colframe=blueish!90!black,title=#1]

\emph{#2}

\end{tcolorbox}
}

%% file: 2-0-abstract.tex
\begin{abstract}
\draft{2}{
    Self-Admitted Technical Debt, or SATD, is a self-admission of technical debt present in a software system. 
    The presence of SATD in software systems negatively affects developers, therefore, managing and addressing SATD is crucial for software engineering. 
    To effectively manage SATD, developers need to estimate its priority and assess the effort required to fix the described technical debt. 
    About a quarter of descriptions of SATD in software systems express some form of negativity or negative emotions when describing technical debt. 
    In this paper, we report on an experiment conducted with 59 respondents to study whether negativity expressed in the description of SATD \textbf{actually} affects the prioritization of SATD. The respondents are a mix of professional developers and students, and in the experiment, we asked participants to prioritize four vignettes: two expressing negativity and two expressing neutral sentiment.
    To ensure the vignettes were realistic, they were based on existing SATD extracted from a dataset. We find that negativity causes between one-third and half of developers to prioritize SATD in which negativity is expressed as having more priority.  
    Developers affected by negativity when prioritizing SATD are twice as likely to increase their estimation of urgency and 1.5 times as likely to increase their estimation of importance and effort for SATD compared to the likelihood of decreasing these prioritization scores. 
    Our findings show how developers actively use negativity in SATD to determine how urgently a particular instance of technical-debt should be addressed.
    However, our study also describes a gap in the actions and belief of developers. 
    Even if 33\% to 50\% use negativity to prioritize SATD, 67\% of developers believe that using negativity as a proxy for priority is unacceptable. 
    Therefore, we would not recommend using negativity as a proxy for priority. However, we also recognize it might be unavoidable that negativity is expressed by developers to describe technical debt. 
    }

    \keywords{Self-Admitted Technical Debt \and Software Engineering \and Sentiment}
    %\subclass{}
\end{abstract}

%% file: 2-1-introduction.tex
\section{Introduction}

Technical Debt is used as a metaphor by developers to describe suboptimal implementations that require future reimplementations to fix the existing implementation~\citep{Cunningham:1992,Tom:2013}.
\draft{1}{Technical debt is pervasive as a large number of developers is familiar, and even affected by, it~\cite{Lim:2012}.
The presence of technical debt in a system is known to have negative effects: it not only makes it more difficult to modify a software project~\cite{Wehaibi:2016},\draft{5}{but also reduces the morale of developers working on a system} where technical debt is present~\cite{Besker:2020}.}

Self-Admitted Technical Debt, or SATD, is a specific category of Technical Debt. 
SATD is characterized by explicit admissions of developers indicating the presence of technical debt~\citep{Potdar:2014}. 
Both technical debt and SATD have been extensively studied, including aspects of SATD such as the automatic identification~\cite{Maldonado:2017,Guo:2021} or the management and removal of SATD~\cite{Maldonado:2017, Zampetti:2018,Zampetti:2020,Tan:2021}.
Specifically, we know that developers use SATD to describe a wide range of technical issues~\cite{Maldonado:2015,Cassee:2022}.
In particular, existing taxonomies of SATD instances show that developers tend to describe functional issues or poor implementation choices.

\draft{5}{Sentiment is a construct used to categorize text into one of the three categories~\cite{Novielli:2023}. Negative sentiment includes texts expressing negative emotions, while text has positive sentiment expressing positive emotions, and a text is considered neutral if it expresses no emotions.} 
The automatic classification of sentiment in software engineering texts has been used to study many different aspects of software engineering ~\cite{Lin:2022}, such as in the code-review process~\cite{Singh:2018}, or on Stack Overflow~\cite{Calefato:2018, Swillus:2023}.
\draft{2}{\draft{5}{\citet{Olsson:2021} found that the presence of technical debt,} or design smells, can cause developers to experience negative emotions.
Furthermore, it has also been found that in roughly 20\% of SATD instances, negativity, or negative emotions, are used to describe the SATD~\cite{Cassee:2022}.}
From the psychological literature, we know that emotionally salient information is more likely to capture attention in the working memory of the brain~\citep{Okon-Singer:2015}.
Similarly, when it comes to issue resolution, previous work has found that there appears to be a link between the expression of positive sentiment and a faster resolution of issues~\cite{Ortu:2015,Sanei:2021}. 
While \citet{Calefato:2018} concluded that Stack Overflow questions that are neutral, or in which no sentiment is expressed, are more likely to receive an answer. 
\draft{1}{Focusing on SATD, 
a quarter of the developers surveyed by~\citet{Cassee:2022} stated that they use negative emotions to describe high-priority SATD, and that they interpret negativity expressed in SATD as a proxy for priority.
Because there is often a gap between respondents' beliefs, and their actions~\cite{Barr:2006}, we want to understand whether expressions of negativity in SATD influence prioritization. }
Therefore, we pose the following research question:

\researchq{sentiment-priority}{Do developers interpret technical debt annotated with negative source-code comments as having a higher priority?}

While there are many different ways to prioritize technical debt~\cite{Lenarduzzi:2021} it is currently unknown whether expressions of negativity influence the perception of the priority of technical debt.
Because of how challenging effort estimation is~\citep{Molokken:2003}, it is important to understand how negativity influences the prioritization of technical debt.
If negativity influences prioritization, it might lead to unintended consequences, as technical debt might be prioritized not because it is important or urgent but because negativity has been used to describe it. 

We use a vignette-based experimental design to address \refrq{rq:sentiment-priority}~\cite{Aguinis:2014}.
By purposefully selecting a realistic set of SATD instances, creating variations of these SATD instances in which negativity is expressed, and assigning them \draft{5}{to participants} we study the effect of negativity on prioritization.
\draft{2}{We sampled respondents from open-source software mailing lists and industrial contacts in the Eindhoven region of the Netherlands.}

\draft{2}{Based on the responses of 59 participants, we conclude that between 30\% to 50\% of developers score the priority of technical debt as higher if negativity is expressed in its description. 
Most importantly, even developers who self-report that they are not influenced by negativity are more likely to increase their estimation of the effort required to fix SATD if negativity is expressed. 
Our results show that developers use SATD not just to describe technical issues but also use negativity in descriptions of SATD as an additional dimension to communicate priority. 
}

%% file: 2-2-methodology.tex
\section{Methodology}
In this section, we describe the experiment conducted to understand whether negativity influences the prioritization of SATD. First, we justify our choice to use controlled experiments. 
Then, we describe the design of the experiment, the instruments used in the experiment, and we explain the method used to analyze the data.\footnote{The code of the statistical modeling is available in a replication package at\\\url{https://doi.org/10.6084/m9.figshare.26863435}.}

\subsection{Choice of Research Method}

In this section, we discuss several potential research methods that could be used to understand the impact of negativity on priority using the framework of \citet{Robillard:2024}, \draft{5}{and we justify our choice to use a controlled experiment.}
\draft{5}{We discuss the potential alternatives to a controlled experiment, our considerations, rationale, and the implications of our decision to use an experiment.}

\paragraph{Alternatives} We first briefly describe the three alternatives we considered: A controlled experiment, a Mining Software Repository (MSR) study, and an interview. 
In a controlled experiment, we would ask participants to prioritize different SATD instances, while for an MSR study, we would analyze whether SATD in which negativity is expressed is removed quicker. 
Finally, in an interview, we would ask participants about their past prioritization practices. 

\paragraph{Considerations:} The most important consideration informing \draft{5}{our choice of methodology is the expectation} that any effect of negativity on the perception of priority might be relatively small.
Previous correlational studies on the impact of sentiment on software engineering report small effect sizes~\citep{Cheruvelil:2019,Calefato:2018, Olsson:2021}.
\draft{5}{Because of these small effect sizes, we require a research method that gives us a high level of control, allowing us to isolate and measure the effect of negativity.}
Furthermore, many factors influence prioritization, as prioritization of Technical Debt is a complex process, and many factors influence how it is prioritized~\cite{Lenarduzzi:2021}. This will also limit the effect of a single factor, such as the expression of negativity, on priority. 
Because of the relatively small effect size, estimating the true effect of negativity on prioritization is more difficult: Many confounding variables might influence the removal of SATD, and in fact, even classifying whether the removal of SATD was purposeful is already challenging~\citep{Zampetti:2018}.
The presence of confounding variables and the noise introduced by potentially inaccurate classifications of removal has made us decide to not use an MSR study. 

The second consideration is the observation that human recollection is not optimal which results in humans misremembering, especially when asked about past events~\citep{Johnson:1993, Raykov:2023}. For an interview study or a survey, in which we ask participants about choices they made in the past related to the prioritization of SATD this might be problematic.
Similarly, while we could ask about current prioritization practices, this still has the downside that our questions would be hypothetical.

\paragraph{Implications:} Because of these two considerations, we opt for controlled experiments as research method to study the prioritization of SATD. 
\draft{2}{The first implication of our choice for controlled experiments is that we have a high level of control~\cite{Stol:2018,Storey:2020}. 
We can use this control to account for as many relevant confounding variables that could also influence prioritization as possible. }
However, the second implication of our choice is reduced realism, as with any experimental study~\citep{Stol:2018}.
Asking respondents to prioritize SATD in an experimental context is not very realistic.
The context in which the respondents usually prioritize SATD is likely not the experimental setup. 
Therefore, the effect we observe in the experiments might not be the effect observed in the field, however, \draft{2}{using an experimental set-up can be expected to allow us to understand with certainty whether any effect exists.}

\subsection{Experimental Design:}
For \refrq{rq:sentiment-priority} we use 
a between-person experimental vignette design, \draft{5}{where each participant is only exposed to one-condition,} according to the guidelines of \citet{Aguinis:2014}. 
To maximize realism of the experiment the vignettes we show to participants are instances of SATD, each containing a snippet of source code containing technical debt and a source code comment describing the technical debt.
We ask participants to score the priority of the vignettes, and by experimentally varying whether negativity is expressed in the source code comment describing the SATD, to investigate whether negativity influences prioritization.

\noindent\emph{Operationalization of Priority:} 
The concept of priority is quite a broad topic; different respondents might interpret the meaning of priority differently. 
Therefore, following existing guidelines~\cite{Aguinis:2014}, we ``split'' the concept of priority into three constructs: \emph{Urgency}, \emph{Importance}, and \emph{Effort}.
Both urgency and importance are common constructs used to operationalize priority~\cite{Middleton:2018,Gavidia-Calderon:2021,Bellotti:2004}, and effort is used to determine the cost of Technical Debt repayment~\cite{Lenarduzzi:2021}. \as{The effort/cost connection seems to assume that cost is relevant for priority. This is kind of reasonable but it should be made explicit, and by preference, supported by empirical evidence. Ultimately, it is not clear how the cost should be used for prioritisation: some developers might prefer to work on easy tasks first, to feel that they are done and over granting them some sense of accomplishment, while others might go headstrong on the most difficult tasks as they might be most rewarding or most important for the project... }

\noindent\emph{Participant recruitment:}
The target population for the experiment are software engineers, and \draft{5}{to increase the response rate, we} do not require any minimum working experience.
Because recruiting of participants for software engineering studies is challenging, we use different channels to recruit participants~\citep{Danilova:2021}.
In particular, we posted the invitation to participate in the experiment on a set of mailing lists previously used to recruit software engineers~\citep{Maldonado:2017, Cassee:2022} and on the social-media pages of the authors, as well as used convenience sampling to invite developers at medium to large software companies in the Eindhoven region of the Netherlands. 
To ensure that we did not burden the mailing lists, we sent out invitations piecemeal, a few a day, and we only posted the call to participate on active mailing lists. 
The Ethical Review Board of Eindhoven University of Technology approved both the experiment and the recruitment strategy.\footnote{Reference: ERB2023MCS17}

\begin{table}

    \centering

    \caption{The questions and response options per question as included in the experiment.}
    \label{tab:survey-questions}
    \small
    \begin{tabular}{p{7cm}p{4cm}}
        \toprule

        \textbf{Question} & \textbf{Response Type} \\ 

        \midrule

        \textbf{Q1} How would you rate the \emph{Urgency} of the listed code-snippet? \emph{In this context we define urgency as whether swift action is required to address the technical debt item. } & \emph{One of: Very low, Low, Medium, High, Very high} \\ \midrule

        \textbf{Q2} How would you rate the \emph{Importance} of the listed code-snippet? \emph{In this context we define importance as the impact of the technical debt item. } & \emph{One of: Very low, Low, Medium, High, Very high} \\ \midrule

        \textbf{Q3} How would you rate the \emph{Effort} required to address the listed code-snippet? \emph{In this context we define effort as the amount of work required to address the technical debt item.}  &\emph{One of: Very low, Low, Medium, High, Very high} \\

        \midrule

        \textbf{Perception (Re-used from \citet{Cassee:2022})} & \textbf{Response type}  \\
        
        \midrule 

        \textbf{Q4} When writing source code, how often do you write source code comments indicating delayed or intended work activities such as TODO, FIXME, hack, workaround, etc.? & \emph{Never, Rarely (Less than once a month), Sometimes (Monthly), Often (Weekly), Very often (Daily)}\\ \midrule

        \textbf{Q5} When authoring comments that describe a problem, how often do you write negative  source-code comments indicating delayed or intended work activities such as TODO, FIXME, hack, workaround, etc.? & \emph{Never, Rarely (Less than once a month), Sometimes (Monthly), Often (Weekly), Very often (Daily)}\\ \midrule

        \textbf{Q6} How often do you come across negative source-code comments indicating delayed or intended work activities such as TODO, FIXME, hack, workaround, etc.? & \emph{Never, Rarely (Less than once a month), Sometimes (Monthly), Often (Weekly), Very often (Daily)} \\  \midrule

        \textbf{Q7} While writing a comment describing an issue in the source-code, 
         I am more likely to write negative comments for issues that I believe are more important. & \emph{Strongly disagree, Disagree, Neutral, Agree, Strongly agree} \\ \midrule

        \textbf{Q8} Writing negative comments to assign extra priority to issues in the source-code is an acceptable practice. & \emph{Strongly disagree, Disagree, Neutral, Agree, Strongly agree} \\ \midrule

        \textbf{Q9} Whenever I come across a source-code comment describing a problem that is particularly negative, I interpret this as a more important issue than a source-code comment describing a problem that is more neutral. & \emph{Strongly disagree, Disagree, Neutral, Agree, Strongly agree} \\ 

        \midrule

        \textbf{Demographics} &  \textbf{Response type} \\

        \midrule

        \textbf{Q10} What is your age? & \emph{Open numerical input} \\ \midrule

        \textbf{Q11 }Which of the following best describes your current employment status? & \emph{One of: ``Employed'', ``Independent contractor, freelancer or self-employed'', `` Student'', ``Not employed'', ``Prefer not to say'', ``Retired''} \\ \midrule

        \textbf{Q12} Which of the following best describes the code you write outside of work? Select all that apply. & \emph{One or more of: ``Contribute to open-source software'', ``Hobby'', ``Freelance/contract work'', ``School or academic work'', ``Bootstrapping a business'', 
        ``I do not write code outside of work''} \\ \midrule

        \textbf{Q13} How many years of programming experience do you have? & \emph{Open numerical input} \\

        \bottomrule

    \end{tabular}
\end{table}

\noindent\emph{Instrument Design}: \cref{tab:survey-questions} contains an overview of the 
questions as included in the survey. 
The instrument is divided into three sections: The first section lists the questions we ask for each vignette. 
To ensure the constructs \emph{Effort}, \emph{Urgency}, and \emph{Importance} were interpreted equally by all participants, we provided the italicized definitions included in \Cref{tab:survey-questions}.
The second section contains a set of questions on the perception of participants about the usage of negativity as a proxy for priority, and the third section contains questions on participants' demographics. 
The questions on \draft{5}{perceptions and} demographics were placed at the end of the survey to prevent them from biasing participants~\cite{Steele:1995}.

The demographic questions included in the survey are related to the respondents' age and working experience.
We record working experience because open and closed source developers are known to manage SATD differently~\citep{Zampetti:2021}. 
Experience can be defined in many different ways~\citep{Siegmund:2014}, and for this experiment, we choose to re-use questions about experience from the Stack Overflow developer survey.\footnote{\url{https://survey.stackoverflow.co/2022#overview}}
We also ask respondents to indicate their age as an open input question, following the recommendations from~\citet{Hughes:2016}. 
We record the age because age tends to influence how people experience emotions~\citep{Yeung:2011}. 
\draft{2}{Finally, we also expect developer's attitude towards the practice of using negativity as a proxy for priority to influence prioritization. 
Therefore, we re-use the closed questions from the study of \citet{Cassee:2022}, in which developers are asked about their perceptions and beliefs about the usage of SATD as a proxy for priority. }
\draft{5}{The questions for each of the four vignettes were mandatory, while all other questions were optional.}

\noindent\emph{Vignette selection:}
For the experiment, the vignettes should be as realistic as possible~\citep{Aguinis:2014}.
Therefore, as vignettes, we select SATD instances from an existing dataset of SATD items. 
The dataset was gathered by \citet{Maldonado:2017}, and we use the version of \citet{Cassee:2022} in which the SATD instances have been categorized based on the type of problem described in the SATD.
We select SATD instances from a single category to minimize the risk that differences between SATD instances influence the results. 
The category we select is \emph{Poor Implementation Choices}, the most common SATD category, in which about 30\% of the SATD instances express negativity. \draft{5}{SATD in this category includes comments like \texttt{``// TODO: define constants for magic numbers''}.}

We select four SATD instances from the dataset; we do not select any more to reduce respondent fatigue and the odds of disengaging.  
Because the comments in the dataset of Cassee~\etal have already been labeled with sentiment polarity, we select two comments that have been labeled as negative and two comments that have been labeled as non-negative. % by Cassee \etal. 
\draft{2}{The dataset of Cassee \etal identifies three different sentiment classes in SATD: negative, non-negative, and mixed. The mixed class, however, occurs in less than 2\% of cases. 
Consequently, we exclude the mixed class and only sample from the negative and non-negative classes. Additionally, due to the negative connotations of technical debt, we expect a low number of positive instance~\cite{Cassee:2022}. Therefore, we exclude these, focusing solely on negative versus non-negative instances.
}

The selection of SATD instances is performed manually: we pay careful attention to ensure that we select instances \draft{2}{that are comprehensible and self-contained, such that the respondents can understand them without becoming fatigued or confused.}
\noindent\emph{Alternative generation:}
For the between-person design of the experiment, we require that each of the four selected vignettes has two variations: a neutral instance and a negative instance. One of the two variations can be randomly assigned to a participant. 
For each of the selected vignettes, we create a manipulated variation expressing a different sentiment polarity than the original. 
The two crucial requirements for these manipulated instances are:
The semantics of the original comment should be preserved, \ie the alternative comment should describe the same problem as the original comment, and the manipulated comment should express the required sentiment polarity. 

To generate manipulated comments we used ChatGPT.\footnote{Version 3.5, accessed in November 2023 at \url{https://chat.openai.com}} 
Through ChatGPT, we aim to reduce the risk that our own perception of what negativity in SATD looks like influences manipulated alternatives we would draft ourselves. 
We manually validate the comment generated by ChatGPT to ensure the manipulated comment meets the previously listed requirements. 
For each SATD instance, we prompt ChatGPT to generate three alternative comments that express a sentiment opposite to the sentiment expressed in the instance. 
\draft{1}{We iteratively refined the prompt used to generate the alternative comments and 
we evaluated the alternatives generated by ChatGPT until all authors were satisfied that each alternative was sufficiently realistic.
This process took several iterations.}
Listed below is the prompt used for the sentiment transfer from negative to neutral:

\begin{displayquote}
    \emph{``Take the following source-code comment, and change the language of the source-code comment to ensure that the resulting source-code comment contains neutral sentiment. Generate three alternative, neutral, source-code comments, but make sure to preserve the original meaning of the comment as much as possible:"}
\end{displayquote}

After finalizing the prompt, three of this paper's authors independently voted to select the best-manipulated variation of the SATD instance from the three alternatives. 
Criteria for voting adhere to the requirements listed above: Did the sentiment transfer work (``Is the comment actually neutral?"), and whether the manipulated comment describes the same problem as the original comment. 
\as{Comment on our experience with sentiment, with SATD etc?} \nc{?}\as{That we are right people to perform the voting task.}
We selected the best alternative comments for each of the four selected SATD instances based on the votes. 
For three manipulated instances, all authors voted for the same alternative; for the fourth instance, two of the three voters preferred one. \as{Is it important to say on which vignette we disagree? Did we see any differences in results for this vignette as opposed to other vignettes?}

\lstset{basicstyle=\small}

\begin{longtblr}[
  caption = {\draft{1}{The vignettes as they were used in the experiment. The source code is matched to either the \emph{Negative} or \emph{Neutral} comment and shown to the participant. The \emph{(M)} indicates that the comment was generated using ChatGPT.} \todo{Use textual elements to highlight differences.}},
  label = {tab:vignettes},
]{
  colspec = {XX[4]},
  rowhead = 1,
  hlines,
} 

        \textbf{Type} & \textbf{Item}  \\

\textbf{Vignette \#1} &
\begin{lstlisting}[language=java]^^J 
public static boolean useThetaStyleImplicitJoins;^^J
public static boolean regressionStyleJoinSuppression;^^J
\end{lstlisting} \\

\emph{Negative} &
\begin{lstlisting}[language=java] ^^J
// USED ONLY FOR REGRESSION TESTING!!!! todo : obviously get rid of all this junk ^^J
\end{lstlisting} \\

    \emph{Neutral (M)} & 
\begin{lstlisting}[language=java] ^^J
// Used only for regression testing! todo: clearly remove all this unnecessary code^^J
\end{lstlisting}

        \\

\textbf{Vignette \#2}
&
\begin{lstlisting}[language=java]^^J
ConstDeclNode constDeclNode = (ConstDeclNode) node;^^J
Node constNode = constDeclNode.getConstNode();^^J
\end{lstlisting} \\

\emph{Neutral} &
\begin{lstlisting}[language=java]^^J
// TODO: callback for value would be more efficient, but unlikely to be a big cost (constants are rarely assigned)^^J
\end{lstlisting} \\ 

\emph{Negative (M)} &
\begin{lstlisting}[language=java]^^J
// TODO: This is frustrating! A callback for value would be more efficient, but unlikely to be a big cost (constants are rarely assigned).^^J
\end{lstlisting} \\

\textbf{Vignette \#3} &
\begin{lstlisting}[language=java]^^J
public class ConstDeclNode extends AssignableNode implements INameNode \{^^J
private final String name;^^J
private final INameNode constNode;^^J
...^^J
\}
\end{lstlisting} \\ 

\emph{Neutral} &
\begin{lstlisting}[language=java]^^J
// FIXME: ConstDecl could be two seperate classes (or done differently since constNode and name never exist at the same time).^^J
\end{lstlisting} \\

\emph{Negative (M)} &
\begin{lstlisting}[language=java]^^J
// FIXME: Ugh, ConstDecl is a mess. It should have been divided into distinct classes (or approached differently) because constNode and name are never in sync.^^J
\end{lstlisting} \\

\textbf{Vignette \#4} &
\begin{lstlisting}[language=java]^^J
if (plot instanceof PiePlot) \{^^J
\ \ \ \ applyToPiePlot((PiePlot) plot);^^J
\}^^J
else if (plot instanceof MultiplePiePlot) \{^^J
\ \ \ \ applyToMultiplePiePlot((MultiplePiePlot) plot);^^J
\}^^J
else if (plot instanceof CategoryPlot) \{^^J
\ \ \ \ applyToCategoryPlot((CategoryPlot) plot);^^J
\}^^J
\end{lstlisting} \\

\emph{Negative} &
\begin{lstlisting}[language=java]^^J
// now handle specific plot types (and yes, I know this is some really ugly code that has to be manually updated any time a new plot type is added - I should have written something much cooler, but I didn't and neither did anyone else).^^J
\end{lstlisting} \\

\emph{Neutral (M)} &
\begin{lstlisting}[language=java]^^J
// Now address specific plot types (and yes, I am aware that this code needs manual updates whenever a new plot type is added - a more advanced implementation could have been developed, but it wasn't, and no other approach was proposed).^^J
\end{lstlisting} 

\end{longtblr}

\Cref{tab:vignettes} contains an overview of the selected vignettes. 
Each vignette combines a short source-code snippet with an accompanying source-code comment that ``admits'' technical debt in the snippet. 
The Type column lists the sentiment polarity expressed by the comment, and the \emph{(M)} tag denotes that the comment on that line has been generated using ChatGPT. 

\noindent\emph{Vignette Assignment:} 
Respondents are assigned to one of the two variations for each vignette. 
However, the assignment of variation to respondents is not fully random. 
When assigning variations, we consider the alternative comments generated by ChatGPT as a blocking factor~\citep{Juristo:2001} \draft{5}{because the comments generated by ChatGPT might influence prioritization scores}. We refer to this blocking varaible as \emph{Manipulation}. 

\begin{table}[]

    \centering

    \caption{The two different flows used for the vignette section of the survey. \emph{Sentiment} is the sentiment expressed in the vignette, while \emph{Manipulated} denotes whether the respondent in the flow sees the original comment or the manipulated one.}
    \label{tab:flows}

    \begin{tabular*}{\linewidth}{@{\extracolsep{\fill}} llcccc}
        \toprule

         &  & $V_1$ & $V_2$ & $V_3$ & $V_4$ \\ \midrule

         \multirow{2}{*}{$Flow\ 1$} & \emph{Sentiment} & \emph{negative} & \emph{neutral} & \emph{negative} & \emph{neutral} \\ 
            & \emph{Manipulated} & \xmark & \xmark & \cmark & \cmark \\ \midrule

            \multirow{2}{*}{$Flow\ 2$} & \emph{Sentiment} & \emph{neutral} & \emph{negative} & \emph{neutral} & \emph{negative} \\ 
            & \emph{Manipulated} & \cmark & \cmark & \xmark & \xmark \\ \bottomrule
    \end{tabular*}

\end{table}

We defined two experimental flows to limit the effect of the manipulation on the outcome. 
\Cref{tab:flows} shows these two flows. 
Observe that we ensure that each respondent sees two neutral and two negative vignettes and sees two manipulated and two original vignettes. 
Participants are randomly assigned to one of the two flows, and within a flow to order of the four vignettes is randomized, such that the impact of the blocking factors is limited~\cite{Atzmueller:2010}. 
\draft{5}{To ensure we do not bias participants, we do not reveal to participants the polarity of the vignette. }

Finally, to validate that the vignettes shown in the experiment were appropriate, and the questions asked were understandable, we piloted the survey with a set of four active software engineers. 
\neil{I'm wondering if all the detailed explanation could be simplified by listing the vignettes. If they are simple enough it would be suitable to explain the approach} \nc{included vignettes and clarity of the section has been improved, we can always drop more information. I'm with you that maybe it is too long, but at the same time I'm afraid that all details are needed to understand the results of the experiment.} \as{Add the vignettes in the appendix?} Based on the feedback of the pilot we made minor text changes to the wording of some questions, and clarified a few concepts a bit better. 
The experiment was hosted using Qualtrics.\footnote{\url{https://www.qualtrics.com/}}

\subsection{Data Analysis}
\label{sec:data-analysis}

\begin{figure}[ht]
    \centering 
        \includegraphics[width=.95 \linewidth, trim={0 4cm 0 4cm},clip]{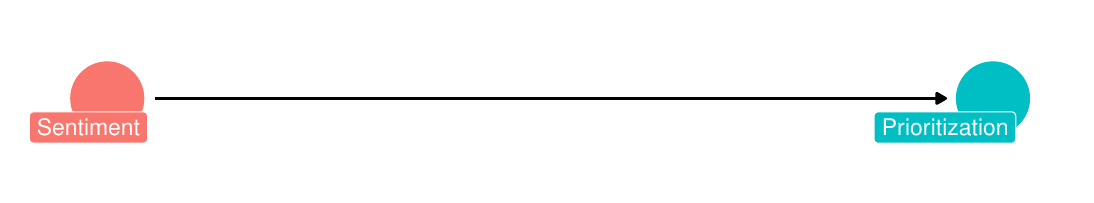}
    \caption{A Directed Acyclic Graph (DAG) illustrating the relation we are analyzing. The color blue denotes the outcome and red the exposure. }
    \label{fig:model_simple}

\end{figure}

\Cref{fig:model_simple} visualizes the relation we study in this manuscript as a Directed Acyclic Graph (DAG).
In a DAG, nodes are used to represent variables, and arrows between nodes represent causal relations between variables~\citep{Elwert:2013}. For instance, \Cref{fig:model_simple} should be interpreted as ``A change in sentiment leads to a change in prioritization''. 

The analysis would be relatively straightforward, assuming that \Cref{fig:model_simple} is the correct model capturing all relevant effects. 
However, while we attempted to control for as many confounding variables in the experimental design, there were confounding factors we could not control for. 
For instance, manipulating the vignettes to transfer sentiment could also have affected the prioritization. 
Therefore, we create a more complete DAG that contains all other variables \draft{5}{in this experimental context} that might influence the prioritization of Technical Debt. 
\as{It is not clear how this more refined DAG has been created. Nathan, I presume that you have done it iteratively and you have consulted Neil?}

\begin{figure}[b]
    \centering 
        \includegraphics[width=.95 \linewidth]{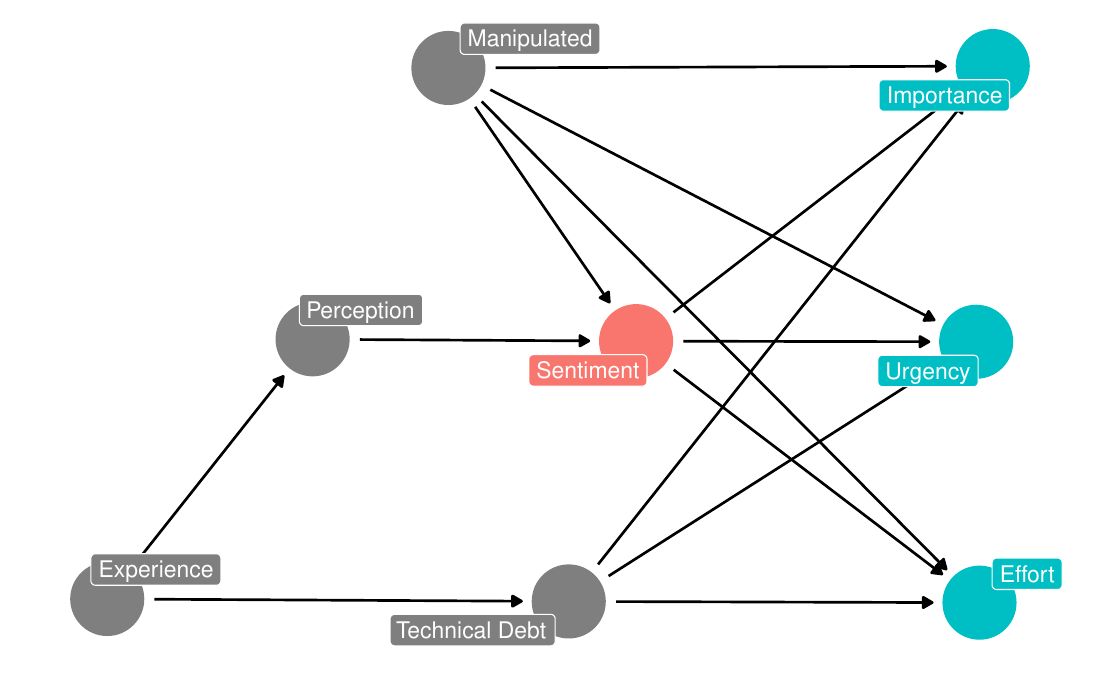}
    \caption{The theoretical model visualized as a DAG. The color blue denotes the outcome, red is the exposure, and grey denotes confounders. }
    \label{fig:model-v1.1}

\end{figure}

\draft{3}{
\cref{fig:model-v1.1} shows the complete mode, including all variables that we could not control for in the design of the experiment. We first explain the construct that each of the variables describes, and then we explain the relations between the nodes. }

\draft{3}{
\begin{itemize}
    \item \textbf{Importance}, \textbf{Urgency} and \textbf{Effort}: This is the node prioritization in \Cref{fig:model_simple} split into one node for each of the Likert-scale questions asked in the experiment. 
    \item \textbf{Sentiment}: Sentiment is the binary exposure variable, recording the sentiment of the vignette shown to the participant. 
    \item \textbf{Technical Debt}: We selected four existing instances of technical debt to use as the basis for the vignettes. While these four instances all belong to the same category of technical debt, there are still variations in the technical debt in each of the vignettes. It is a categorical variable, recording which of the four vignettes the participant responded to. 
    \item \textbf{Manipulated}: Each vignette in the experiment has two variations: The unmodified, existing instance of technical debt and one variation for which the alternative comment was generated using ChatGPT. Manipulated is a binary variable indicating whether a participant was looking at a variation generated by ChatGPT.  
    \item \textbf{Perception}: From \citet{Cassee:2022} we know that developers have different perceptions concerning the usage of negativity as a proxy for priority. Whether a participant agrees, has no opinion, or disagrees with using negativity as a proxy for priority is indicated by this variable as a trinary variable. 
    \item \textbf{Experience}: Experience captures the participant's years of professional working experience.
\end{itemize}
}

\draft{3}{
We expect that the prioritization, measured as \emph{Importance}, \emph{Urgency} and \emph{Effort}, is directly influenced by the \emph{Sentiment} in the vignette. Because of the results of the survey of \citet{Cassee:2022}.
Secondly, the \emph{Technical Debt} shown to participants is different in each vignette; therefore, we expect that each item of technical debt will receive a different prioritization score. 
Finally, we expect whether we \emph{Manipulated} the comment to influence both prioritization and sentiment. While we did our best to ensure the comments generated using ChatGPT were as realistic as possible, the manipulated comments might still contain language that influences the prioritization given by participants. Additionally, comments generated by ChatGPT could also express negativity differently than human-authored comments, and therefore, whether a comment is \emph{Manipulated} could affect how \emph{Sentiment} influences prioritization.
}

\draft{3}{
Whether a participant believes negativity should be used to prioritize technical debt (\emph{Perception}) will influence whether a participant uses \emph{Sentiment} to prioritize. We expect the \emph{Experience} of participants to influence both how they prioritize \emph{Technical Debt}, and their \emph{Perception} to using negativity as a proxy for priority~\citep{Yeung:2011,Zampetti:2021}. \ie more experienced participants will have seen more technical debt in their career and therefore prioritize specific instances of technical debt. Similarly, for \emph{Perception}, a more experienced candidate could be more likely to think using negativity is an acceptable prioritization mechanism. 
}

Given the DAG shown in \Cref{fig:model-v1.1}, understanding the effect of negativity on prioritization (measured as \emph{Effort}, \emph{Urgency}, and \emph{Importance}) requires building models that adjust for the other variables in the DAG. 
\draft{3}{Specifically, understanding the total effect of \emph{Sentiment} on \emph{Prioritization} requires adjusting for \emph{Manipulated}, \emph{Perception} and \emph{Experience}.
To accurately estimate the effect of sentiment on prioritization, we use Bayesian statistics~\citep{Gelman:2013}.
In frequentist statistics, model outcomes are often single-point estimates of effects and significance values. 
One benefit of Bayesian statistics is the fact that model output is not just a single-point estimate.
Instead, because we use Bayesian statistics, we can better account for any uncertainty in the data~\cite{Furia:2022, McElreath:2018, Kruschke:2018}.   
By following existing guidelines \citep{Gelman:2020:Workflow, Furia:2022} and the examples of the applications of Bayesian statistics to software engineering data \citep{Torkar:2022,Ghorbani:2023} we ensure we can estimate the effect of negativity on prioritization.}

In the remainder of this section, we first describe the Bayesian models we fit to the data. 
Then, we explain how we use the models to understand the effect of negative sentiment on the prioritization of SATD.

\begin{equation}
  \tag{Model 1}
  \begin{split}
  Outcome & \sim \text{Ordered-logit}(\phi,\ \kappa) \\
  \phi & = \alpha_{\mbox{\sl effects}[\textcolor{red}{{Sentiment}},\ \textcolor{red}{{Perception}}]} + \alpha_{\mbox{\sl manipulated}[\textcolor{red}{{Manipulated}}]} \\
  &\ \  + \alpha_{\mbox{\sl experience}} * \textcolor{red}{\mbox{\sl Experience}}   \\
  \alpha_{\mbox{\sl effects}} & \sim \text{Normal}(0, 0.5)\  \\
  \alpha_{\mbox{\sl manipulated}} & \sim \text{Normal}(0, 0.5)\ \\\
  \alpha_{\mbox{\sl experience}} & \sim \text{Normal}(0, 0.5)\  \\
  \kappa & \sim \text{Normal}(0, 2)\ \\
  \text{Where} & \ Sentiment,\ Manipulated \in \{\text{True},\ \text{False}\} \\ 
  &\ Perception \in \{\text{Disagree},\ \text{No Opinion},\ \text{Agree}\} \\
  &\ \mbox{\emph{Experience}} \in \mathbb{Q}
  \end{split}
  \label{eqn:model_adjust}
\end{equation}

\noindent \emph{Model:}
Based on the causal graph of \Cref{fig:model-v1.1}, we should adjust for three variables: \emph{\textcolor{red}{Perception}}, \emph{\textcolor{red}{Manipulated}}, and \emph{\textcolor{red}{Experience}}.
\ref{eqn:model_adjust} shows the model we use, where \emph{Outcome} is the score given by a participant to the vignette. Because we have three outcome constructs: \emph{Effort}, \emph{Importance}, \emph{Urgency}, we fit one model per construct.
The first line shows the choice of the \emph{likelihood function}, an Ordered-logit. 
lines 2--3 show the choice of model parameters, indicated by the prefix $\alpha$, lines 4--8 show the choice of priors, and lines 9--11 show the allowed values for the input variables.  \as{Maybe add ``We explain these decisions below'' or something similar? Otherwise, there is a voice in my head shouting ``why, why, why'' after each sentence.}

Because the outcomes are responses on a Likert scale. \ie the responses are on an ordinal scale, we use an Ordered Logit as a likelihood function as recommended by \cite{McElreath:2018}.
Ordered-logit has two arguments: $\phi$, a model-term consisting of all of the parameters (denoted by $\alpha$) and $\kappa$, the cutpoints, used to model the log-cumulative-odds for the Likert responses. 
The model itself has four input variables (typeset in red and italics): Whether the vignette shown to the developer was negative (\textcolor{red}{$Sentiment$}), whether the vignette was manipulated (\textcolor{red}{$Manipulated$}), what the perception of the respondent was (\textcolor{red}{$Perception$}), and the experience of the respondent (\textcolor{red}{$Experience$}).
Negative and manipulated are binary variables, perception is a trinary variable based on the respondent's answer to Q9\nc{Consider explaining why Q9}, and age is a natural number.  
$\mbox{\sl effects}$ 
is a 2 by 3 matrix with a model parameter for each combination of \textcolor{red}{$Sentiment$} and \textcolor{red}{$Perception$}. Manipulated is a vector of length 2, for each value of \textcolor{red}{$Manipulated$}, and experience, \textcolor{red}{$Experience$}, is a single parameter. 

\draft{3}{The priors we select are all uninformative, and we do not associate any positive or negative effects to them. 
We opted for uninformative priors as we did not want to bias the models. Selecting priors that expect sentiment to increase prioritization scores could potentially bias our results, and there is no literature that allows us to pick more informative priors. } 

\noindent \emph{Estimating Effects:} We use the three models, fit for each of the three outcome variables, to quantify the impact of negativity on prioritization. 
First, we plot the distribution of the model parameters related to sentiment, the so-called posterior, as density plots. 
\draft{1}{The posterior plots show the effect on the outcome variable that the model associates with a change from neutral to negative sentiment and are commonly used to interpret results of Bayesian models~\cite{McElreath:2018}.
We also use the posterior to compute the \emph{Evidence Ratio} of the effect of negativity and interpret the evidence ratio according to \citet{Stefan:2019}.} \draft{5}{Where an evidence ratio larger than one is considered as evidence in favor of a correlation, and an evidence ratio smaller than one is considered as evidence for an inverse correlation.}

\draft{1}{Through the posteriors we can understand whether negativity influences prioritization, however, it does not allow us to quantify how often negativity leads to a different prioritization score. 
Therefore, we also use the fitted models to simulate the effect of negativity \ie \emph{``What are the odds that changing sentiment from neutral to negative increases the prioritization score?''}} 
To do this, we manually fix the input variables (\textcolor{red}{$Sentiment$}, \textcolor{red}{$Manipulated$}, \textcolor{red}{$Perception$}) to obtain two simulated distributions of prioritization scores for a SATD instance: a set of prioritization scores for an SATD instance with neutral sentiment, and SATD instance with negative sentiment. 
We can quantify how negativity influences the outcome variable by computing the contrast between these two distributions. 
Because \ref{eqn:model_adjust} has as input the perception of the respondent, whether the vignette has been manipulated, and the experience of the respondent we have to fix values for all three. 
As we are not interested in a scenario in which the vignettes are manipulated, we fix \textcolor{red}{$Manipulated$} to false. Similarly, we set \textcolor{red}{$Experience$} to the mean experience value of the respondents.
Finally, we can not fix the value of \textcolor{red}{$Perception$} to one particular value for perception. Therefore, we compute a contrast distribution for each value of \textcolor{red}{$Perception$} (Disagree, Indifferent, Agree).

\begin{figure}[t]
\begin{subfigure}{.32\textwidth}
  \centering
  \includegraphics[width=1\linewidth]{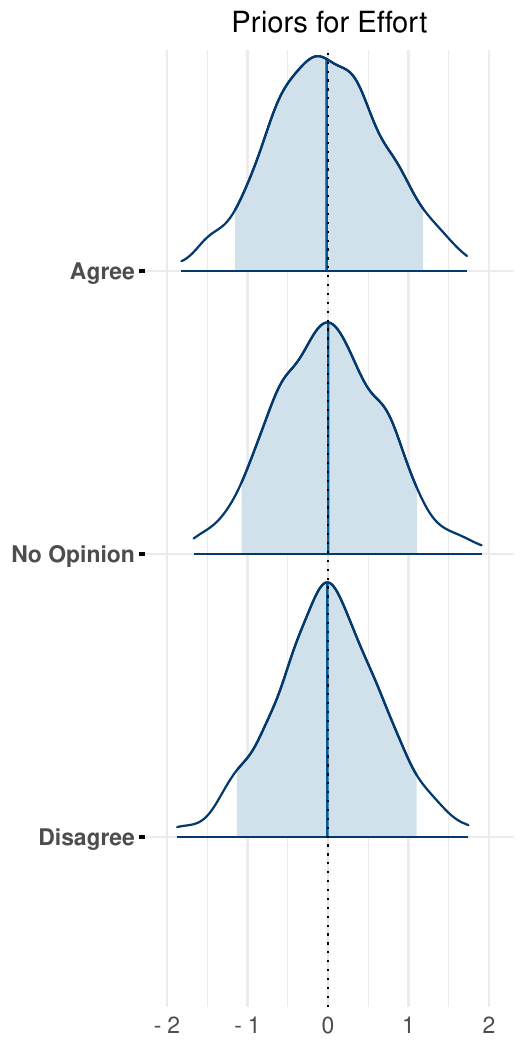}
  \caption{}
\end{subfigure}
\begin{subfigure}{.32\textwidth}
  \centering
  \includegraphics[width=1\linewidth]{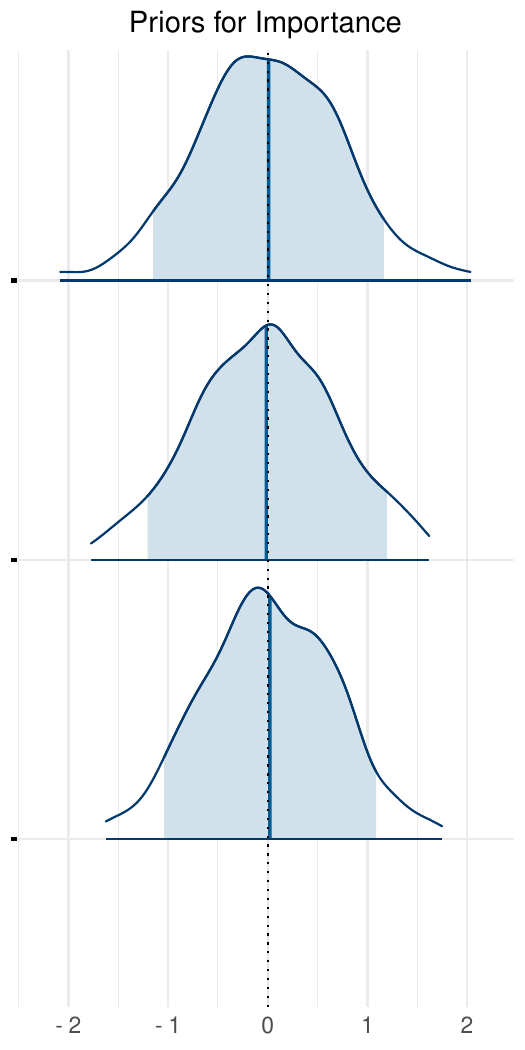}
  \caption{}
\end{subfigure}
\begin{subfigure}{.32\textwidth}
  \centering
  \includegraphics[width=1\linewidth]{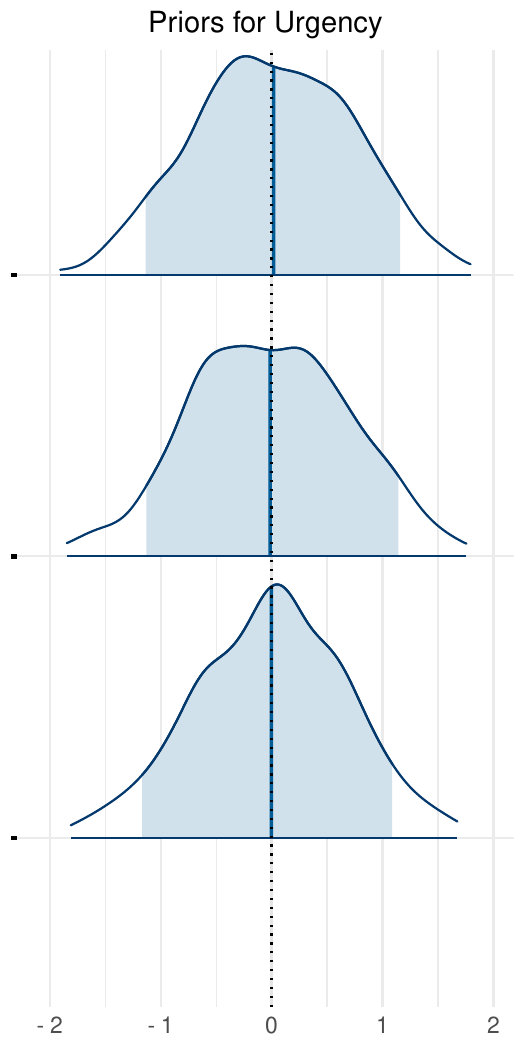}
  \caption{}
\end{subfigure}
\caption{Distributions of the priors for each of the outcome variables. }
\label{fig:priors}
\end{figure}

\begin{fancylongtblr}
[
caption = {The odds ratio between the odds of the priority score increasing compared to the odds of the score decreasing if sentiment changes from neutral to negative.},
label = {tab:priors-odds-ratio},
]
{
colspec = {l X[3,r] X[3,r] X[3,r]}, 
}

Perception & Effort & Urgency  &  Importance   \\

Agree & 1.02 & 1.00 & 0.99 \\
No Opinion & 1.00 & 0.99 & 1.00 \\
Disagree & 1.00 & 0.98 & 1.00 \\
\end{fancylongtblr}

\paragraph{Prior predictive checks:}
\draft{3}{To ensure our choice of priors does not bias our results, we perform prior predictive checks.
Specifically, we sample from the model using only the priors to verify that the outcomes of the model are realistic. \ie the model only generates responses within the range of 1 -- 5. Secondly, we want to verify that the responses of the model are unbiased, \ie the priors do not associate any effect of negativity on prioritization.
To verify the outcomes are unbiased, we compute the odds of the prioritization increasing if negativity is expressed, as described in \emph{estimating effects}. 
\Cref{fig:priors} plots the distribution of the priors for the effect of negativity on the estimation of the outcome variables. 
The distribution for each of the priors is centered around zero, showing that the priors do not favor any specific outcome. Additionally, if the priors are unbiased, we expect the odds ratios to be close to zero. This is the case, as can be seen in \Cref{tab:priors-odds-ratio}.}

\paragraph{Perception and Demographics} Respondents' answers to the closed questions and demographics are visualized. 
As the questions on Perception are taken verbatim from the study of \citet{Cassee:2022} the results obtained in this experiment are compared to those of the original study. 
Meanwhile, the answers to questions on demographics taken from the Stack Overflow survey will be compared to the most recent results of the Stack Overflow survey. 
\as{I understand where this paragraph comes from but after an extensive discussion of Bayesian modeling, this paragraph somehow feels anticlimactic. Not sure what can be done here though.}

%% file: 2-3-results.tex
\section{Results}

In total, we received 75 full and partial responses to the experiment. Based on a manual check, we removed one response because the values provided by the respondent for age and experience were unrealistic.
Because we also accepted partial responses and because the instrument contained optional questions, we discuss the number of relevant and full responses included per question.

\subsection{Demographics}

\begin{figure}[t]
\begin{subfigure}{.48\textwidth}
    \centering
    \includegraphics[width=1\linewidth]{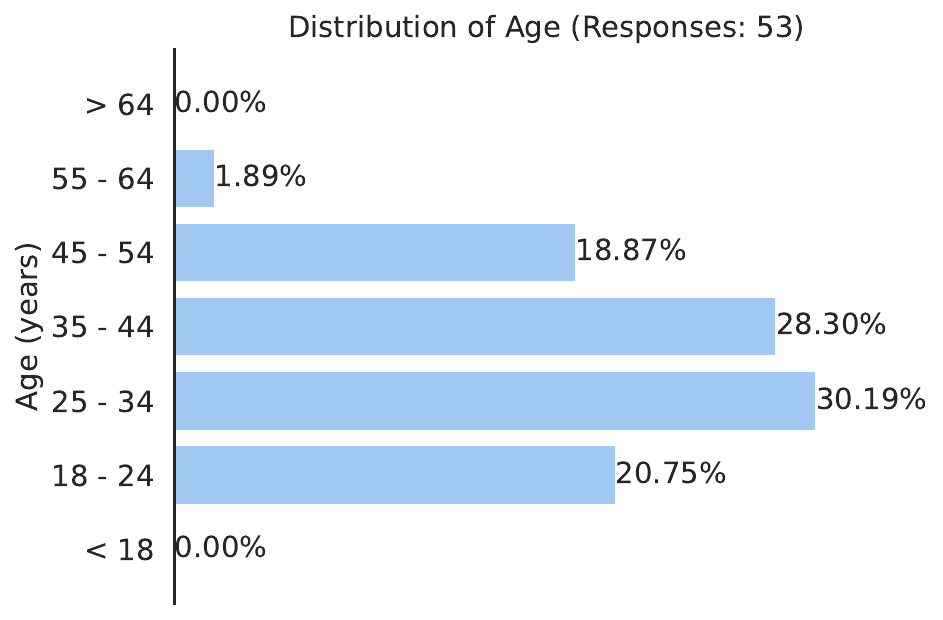}
    \caption{}
    \label{fig:hist-age}    
\end{subfigure}
\begin{subfigure}{.48\textwidth}
    \centering
    \includegraphics[width=1\linewidth]{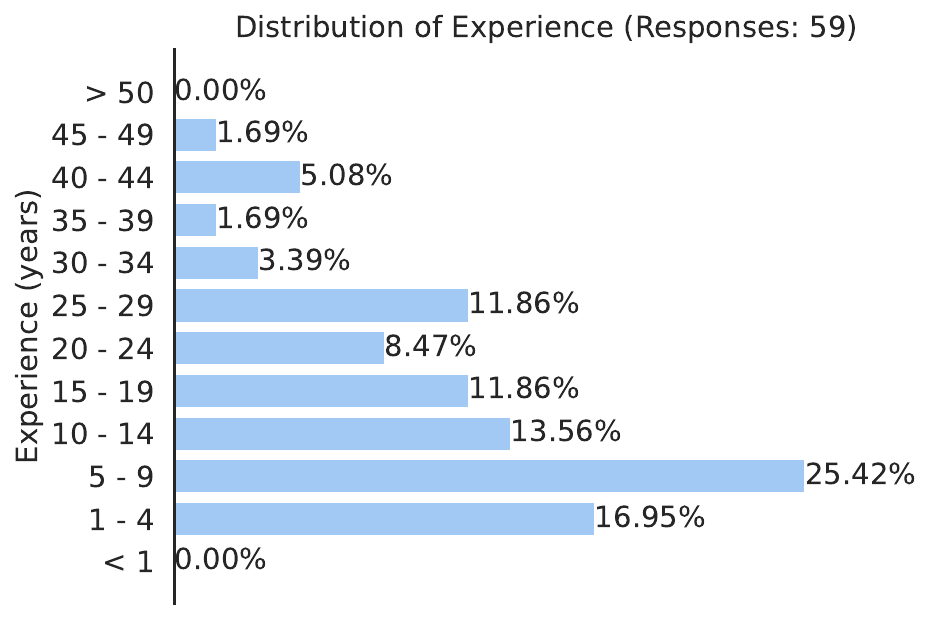}
    \caption{}
    \label{fig:hist-experience}    
\end{subfigure}
\caption{Histograms for respondent age and experience.}
\label{fig:hist-demographics}
\end{figure}

\Cref{fig:hist-demographics} shows the respondent age and experience distributions. The bins for both plots are identical to the bins of the Stack Overflow developer survey.\footnote{\url{https://survey.stackoverflow.co/2023/#overview}} 
\draft{1}{We compare the demographics to the Stack Overflow developer survey because the Stack Overflow survey is one of the largest surveys of developers, with almost 90,000 respondents for 2023. }
For age, our experiment's population appears to be a bit older than the responses to the Stack Overflow developer survey. Notably, we saw no respondents younger than 18 or older than 64. 
Experience-wise, the distribution of respondents is quite similar to that of the Stack Overflow survey. 
\draft{2}{Most importantly, there are no large differences between our respondents and the respondents to the Stack Overflow survey.} Therefore, we conclude, \draft{2}{based on the distributions of age and experience, that our respondents are a good representation of the general developer population for these two characteristics.}

\begin{fancylongtblr}
    [
    caption = {The self-described employment status and the self-described coding outside of respondents' work.},
    label = {tab:demographics-employment-hobby}
    ]
    {
    colspec = {X[3] r r  X[3] r r},
    row{6} = {font = \bfseries, abovesep+=2pt, belowsep+=2pt},
    hline{6-7} = {1pt},
    vline{4} = {1pt},
    baseline = {b}
    }

Employment Status & \# & \% & Coding outside of Work & \# & \% \\ 

Employed full-time & 45 & 76.27\% & Personal projects & 20 & 64.52\% \\
Student, full-time & 8 & 13.56\% & I do not write code outside of work & 7 & 22.58\% \\
Self-employed & 4 & 6.78\% & School or Academic & 3 & 9.68\% \\
Employed part-time & 2 & 3.39\% & Open-source & 1 & 3.23\% \\
Total & 59 & 100.00\% & Total & 31 & 100.00\% \\
\end{fancylongtblr}

\Cref{tab:demographics-employment-hobby} shows respondents' employment status and the coding respondents do outside of work. 
A large majority of the respondents are professional developers who are employed full-time.

\subsection{Negativity's Effect on Prioritization}

For the Bayesian models, we can only use responses for which all questions used in the model have received a response, particularly the demographic question on respondent experience (Q13).  
This requirement leaves 59 valid responses for the experiment on which the models have been fit. 

\begin{figure}[b]
\begin{subfigure}{.32\textwidth}
  \centering
  \includegraphics[width=1\linewidth]{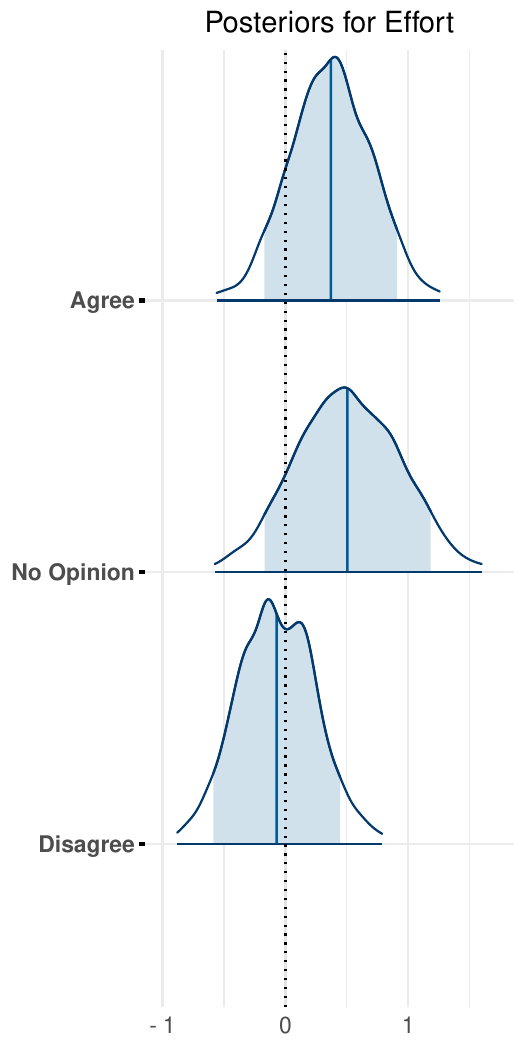}
  \caption{}
\end{subfigure}
\begin{subfigure}{.32\textwidth}
  \centering
  \includegraphics[width=1\linewidth]{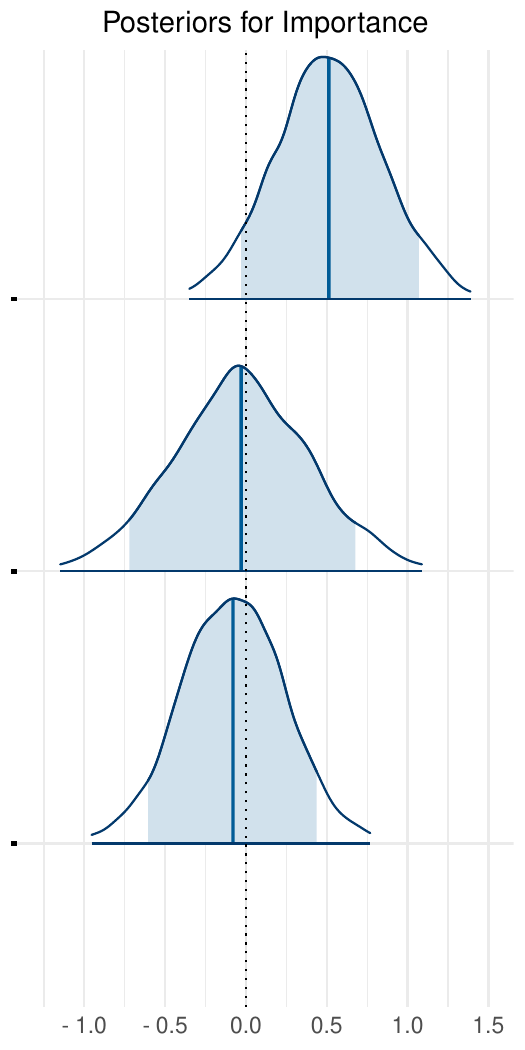}
  \caption{}
\end{subfigure}
\begin{subfigure}{.32\textwidth}
  \centering
  \includegraphics[width=1\linewidth]{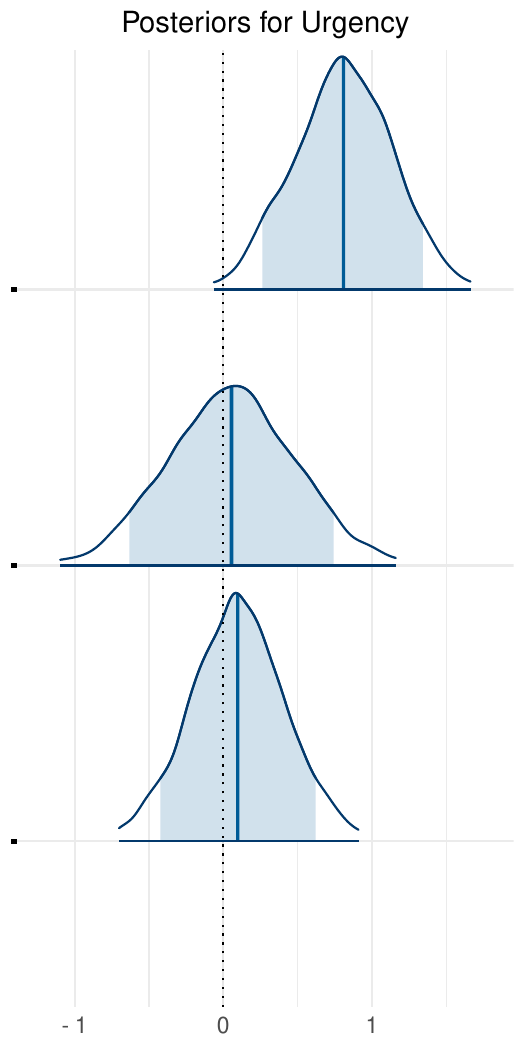}
  \caption{}
\end{subfigure}
\caption{Distributions of the posterior for each of the outcome variables. }
\label{fig:posteriors}
\end{figure}

\draft{1}{\Cref{fig:posteriors} visualizes the density distribution for the posteriors related to the effect of sentiment on the outcome. 
In other words, this figure shows the effect the model associates with a change in sentiment from neutral to negative.
A positive value indicates that \draft{1}{a change from neutral to negative increases} the prioritization score. 
As can be observed in \Cref{fig:posteriors}, the distributions of the priors are generally quite wide, and they overlap with the dashed line indicating zero. 
} 
However, this uncertainty is not unexpected or uncommon in Bayesian analysis, especially in studies with smaller samples~\cite{Ghorbani:2023, Frattini:2024}.

\begin{fancylongtblr}[
        caption = {Evidence ratio table for the hypothesis that negativity increases the prioritization score for each of the outcome variables.},
        label = {tab:evid-ratio}
    ]
    {
        colspec = {X X X X},
    }

Perception & Effort & Urgency & Importance \\ 
Agree & \textbf{Moderate for} & \textbf{{Strong for}} & \textbf{{Strong for}} \\ 
No Opinion & \textbf{Moderate for} & Anecdotal for & Anecdotal against \\ 
Disagree & Anecdotal against & Anecdotal for & Anecdotal against \\ 

\end{fancylongtblr}

\draft{1}{Because we adjusted the models for the \emph{Perception} of the participants, the effect of sentiment on the outcome can vary for each of the three levels of \emph{Perception}. 
The labels \emph{Disagree}, \emph{No Opinion}, \emph{Agree} therefore show the effect of negative sentiment on prioritization based on whether the participants believe that negativity should be interpreted as a signal that the SATD has a higher priority. 
The posteriors indicate that participants who agreed with the use of negativity as a proxy for priority were also more likely to assign a higher priority score, for all of the three measured variables. 
More interestingly, participants who indicated they had no opinion on the use of negativity as a proxy for priority increase the score for effort. Finally, for the other groups and outcomes we see that the models do not associate any impact of negativity, as the means all appear to be centered around zero.
\Cref{tab:evid-ratio} lists an interpretation of the Bayes factor, or the strength of the evidence, of the hypothesis that negativity increases the prioritization. 
We interpret the values for the Bayes factor according to \citet{Stefan:2019}.}

\begin{fancylongtblr}
[
caption = {The odds ratio between the odds of the priority score increasing compared to the odds of the score decreasing if sentiment changes from neutral to negative. Bold font indicates whether the evidence ratio supports the hypothesis that negativity increases prioritization (\Cref{tab:evid-ratio}).},
label = {tab:odds-ratio},
]
{
colspec = {l X[3,r] X[3,r] X[3,r]}, 
cell{2}{2} = {font = \bfseries},
cell{2}{3} = {font = \bfseries},
cell{2}{4} = {font = \bfseries},
cell{3}{2} = {font = \bfseries},
}

Perception & Effort & Urgency  &  Importance   \\

Agree & 1.40 & 1.95 & 1.50 \\
No Opinion & 1.56 & 1.07 & 0.98 \\
Disagree & 0.95 & 1.10 & 0.94 \\
\end{fancylongtblr}

However, quantifying the actual impact that negativity has \eg answering the question: \emph{``Does negativity make it more likely that SATD receives a higher priority score?''} is not possible based only on the plotted posteriors. 
Therefore, we also used the models to simulate, per outcome variable, the difference in priority scores for SATD expressing either neutral or negative sentiment.
This results in a set of probabilities: How likely is it that negativity results in a higher priority score? How likely is it that negativity results in a lower priority score? 
\Cref{tab:odds-ratio} shows the ratio between these two odds for each outcome variable and for each value of \emph{Perception}. 
From this table, we conclude that \draft{5}{37\%} of respondents who interpret negativity as a signal for importance (i.e., ``Agree'') are also more likely to assign higher priority scores for SATD expressing negative sentiment. 
In this case, the consistency between respondents' beliefs and actions is noteworthy, as the value-action gap is a well-documented bias~\cite{Barr:2006}. 
Secondly, \Cref{tab:odds-ratio} also shows that the perception of Urgency (OR: 1.95) is more likely to be influenced by negative sentiment than Effort (OR: 1.40) and Importance (OR: 1.50). 
Finally, \draft{5}{20\%} of respondents who had no opinion on whether they interpreted negativity as a signal of importance are likelier to assign a higher score for effort when negativity is expressed.

\implication{Findings -- Negativity}{Up to 57\% of developers estimate the priority of self-admitted technical debt as higher when negativity is used to describe it. These developers are between 1.4 and 2.0 times more likely to increase, instead of decrease, their prioritization scores for self-admitted technical debt expressing negativity.    }

\subsection{Perceptions and Self-Reported Behavior}

\begin{figure}[t]
  \centering
  \includegraphics[width=.95\linewidth]{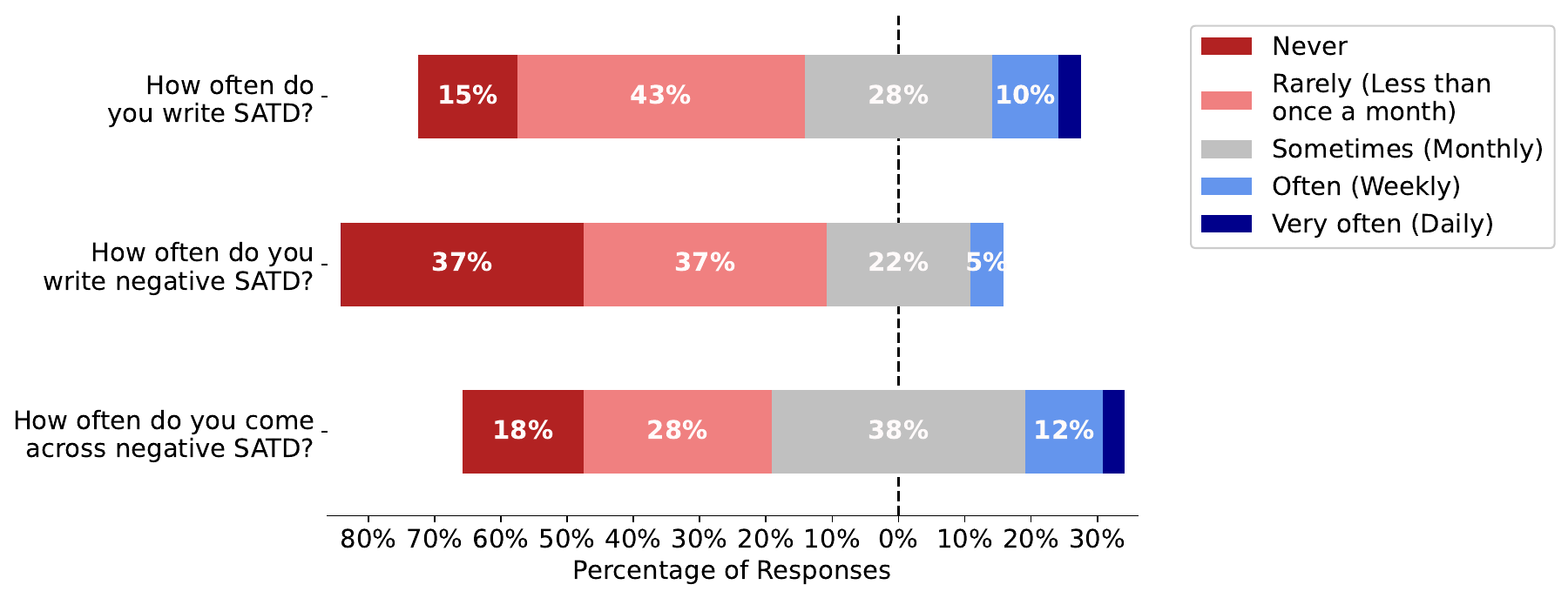}
\caption{Responses to survey questions Q4, Q5, Q6.}
\label{fig:likert-frequency}
\end{figure}

\Cref{fig:likert-frequency} shows the responses to the questions asking participants about how often they write or come across SATD that expresses negative sentiment. 
These results indicate that almost half of the developers encounter SATD expressing negative sentiment monthly or even more often. 
However, most developers, almost 75\%, never, or rarely, write any SATD expressing negative sentiment.
The distributions observed in this study are similar to the study of \citet{Cassee:2022}. 
The most important difference between this study and the findings of Cassee~\etal is that respondents to the experiment tend to come across SATD expressing negativity less frequently.
\draft{2}{This could be explained by the fact that we used convenience sampling through our industrial contacts for this experiment, and because industrial developers annotate SATD differently~\cite{Zampetti:2021}.}

\begin{figure}[t]
  \centering
  \includegraphics[width=.95\linewidth]{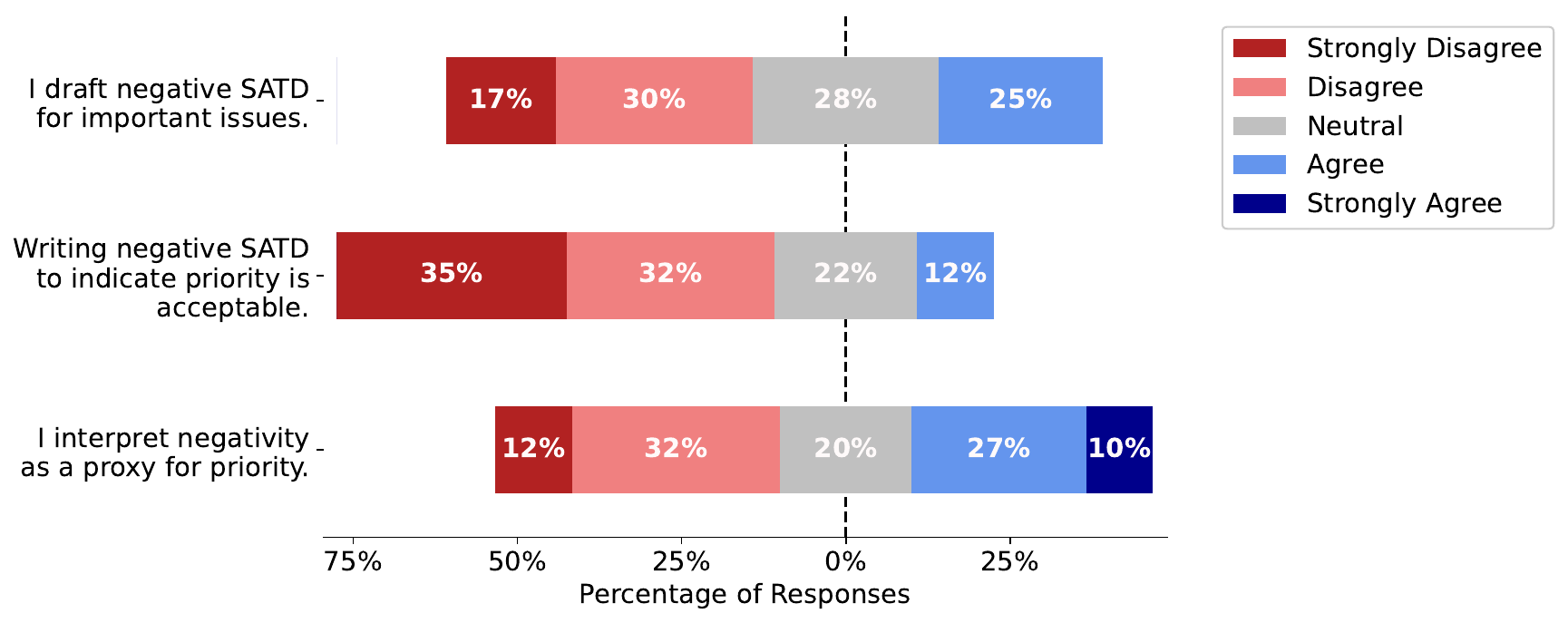}
\caption{Responses to survey questions Q7, Q8, Q9.}
\label{fig:likert-behavior}
\end{figure}

\Cref{fig:likert-behavior} outlines the results to the questions on respondents perceptions. 
Compared to \citet{Cassee:2022} the proportion of respondents indicating that they strongly disagree with the statement that they interpret negativity as a proxy for priority is smaller. %\as{the same question as above.}
For the other two statements the results for this experiment are similar to those of Cassee~\etal. 

From the combination of \Cref{fig:likert-frequency} and \Cref{fig:likert-behavior}
we generally confirm the findings of \citet{Cassee:2022}. 
A large majority of developers disagree that writing SATDs that express negative sentiment to signal priority is acceptable. However, even if most developers believe this, they still express negativity in SATDs to indicate priority or state that they interpret negativity as a proxy for priority. \as{this discussion seems to be OK but it glosses over the differences between the studies.}

\implication{Findings -- Perceptions}{Between 25\% to 50\% of developers draft and/or encounter self-admitted technical debt expressing negativity. Meanwhile, 67\% of developers state that using negativity as a proxy for priority is unacceptable.   }

%% file: 2-4-discussion.tex
\section{Discussion}

Our study shows that developers use source-code comments to communicate not only a description of technical debt but also priority.
From previous work, we know that source code comments have always been used for various purposes, including the description of technical-debt~\cite{Pascarella:2017}. 
Furthermore, descriptions of technical debt in source-code comments cover a wide range of technical issues~\cite{Maldonado:2017,Cassee:2022}.
Based on the results of our experiment, we add to the existing body of knowledge on the role of source-code comments in addressing technical debt. Specifically, we find that negativity in these descriptions results in the same technical issue being prioritized differently. 
Therefore, we conclude that developers use source code comments not just to describe technical issues but also use negativity to communicate priority.
Negativity can be seen as a hidden message or additional dimension, as a SATD comment's primary purpose is to describe the technical debt. This study and the results of the previous study~\cite{Cassee:2022} show that developers purposefully express negativity to emphasize priority and interpret negativity as a proxy for priority.

\paragraph{Does this mean developers should use negativity to describe technical-debt they think is important?} 
For individual developers, using negative expressions to annotate the technical debt they are working on might be an effective strategy. 
It could result in the issue they care about being prioritized and, therefore, fixed before other technical debt is addressed. 
However, from a team perspective, this is a potentially problematic strategy. 
Software projects are often characterized by only having limited capacity available to work on technical debt~\cite{Zampetti:2021}, and developers using negativity to describe technical debt lay claim to more of this capacity.
Secondly, whether developers are influenced by negativity is conditional to their perceptions, so not everyone in a team will interpret the priority similarly. 
Additionally, two-thirds of developers believe that using negativity as a proxy for priority is unacceptable.
Finally, technical debt demoralizes developers~\cite{Besker:2020}, and expressions of negativity could be particularly demoralizing; therefore, we would not recommend developers use negativity as a proxy for priority \draft{5}{as there are alternative prioritization methods.} 
\draft{5}{From \citet{Cassee:2022}, we know developers prefer the use of other channels to communicate the priority of SATD, such as issue trackers or work-management systems. }

\paragraph{Does this mean developers should stop expressing negativity?}
Emotions and sentiment are a part of everyday life~\cite{Kleef:2009}. 
It has long been found that developers express emotions in many different software engineering activities~\cite{Calefato:2018,Mantyla:2016,Lin:2022}, 
and technical debt appears to be able to trigger negative emotions in software engineers~\cite{Olsson:2021}. 
This makes it unrealistic to expect developers to stop expressing their emotions or opinions completely. 
Instead, this study shows how negativity can affect the management of technical debt, \draft{5}{where prioritization is one of the many different management activities of technical debt~\cite{Li:2015}.} 
However, we believe it is important to emphasize that any negative emotions expressed are not directed towards developers, as they sometimes are~\cite{Gachechiladze:2017}. 
Especially as negativity directed towards other developers could be considered toxic, with far-reaching consequences~\cite{Miller:2022, Raman:2020}.

\paragraph{Generalizability \& Future work:}
We expect the findings of this study to apply to technical debt in general and not just to self-admitted technical debt described in source-code comments. 
\draft{5}{Because of the high level of control in the experiments and the experienced participants, we expect the observed effect of negativity on prioritization to translate to technical debt described on other platforms.} 
Notably, this includes places like issue-trackers, as these are places where technical debt is often described~\cite{Cassee:2022,Zampetti:2021, Xavier:2022}.

Through the experiment, we conclude that negativity affects prioritization, and we describe how developers use negativity in source-code comments to communicate priority. 
However, a consequence of the high control of the experiments is the reduced realism. 
Therefore, there is still an opportunity for future work to understand how negativity in technical debt influences prioritization in the ``field''~\cite{Storey:2020}. 
Research methods or strategies like ethnographic observations can describe the prioritization of technical debt in software projects, as opposed to the contrived setting of the experiments, like the work of \citet{Aranda:2009} on bugs. 
\draft{5}{Or, research methods such as mining software repositories can be used to study whether SATD in which negativity is expressed is removed quicker than neutral SATD.}

\draft{5}{Secondly, in this study, we only focus on the effect of negativity, and we consider the effect of positivity to be out of scope. This is done purposefully, as previous work has found that positivity rarely occurs in SATD~\citep{Cassee:2022}. Therefore, understanding its effect on prioritization is not as relevant. However, prior literature has described other software engineering activities where developers express positivity in their communication~\cite{Calefato:2018,Asri:2019,Cheruvelil:2019}. We believe that these activities (change requests, asking questions on SO) are better suited to study the effect of positivity.}

%% file: 2-5-threats.tex
\section{Threats to Validity}

Notwithstanding our effort to ensure our results were valid, factors outside of our control might still have influenced the results. 
To discuss these threats to validity and our mitigations, we use the framework of \citet{Wohlin:2012}, as the research method of this study is an experiment. 

\paragraph{Internal Validity:}
We carefully designed the experiments to reduce the effect of confounders as much as possible, for instance, by randomizing the order of vignettes, using existing instances of SATD, and by making sure the instances were understandable.
However, to create the vignettes we manipulated descriptions to create either a negative or a neutral counterpart. 
We took several precautions to reduce the bias of the manipulation of the obtained results. 
These include ensuring a balance by creating two negative and two neutral manipulated descriptions, generating several manipulated descriptions, and voting for the best one. 
Furthermore, we added manipulation as a parameter to the model to adjust for any remaining bias.

Secondly, we did not state the exact purpose of the experiments to prevent the respondents from guessing the experiment's hypothesis and responding in line with their \emph{attitude} towards the hypothesis. 
Similarly, the vignettes did not indicate whether participants viewed the neutral or the negative instance.
However, there is still a chance that some respondents might have recognized the purpose of the experiments after scoring two or more vignettes and were, therefore, influenced by their attitudes. 

\draft{5}{Finally, there is a risk that the wording of the perception-related questions might have biased participants. However, re-using the exact wording from the survey administered by \citet{Cassee:2022} allows us to compare our results to those of Cassee \etal.}

\paragraph{External Validity:}
The respondents sampled for the experiment were diverse with respect to age and working experience, \draft{5}{with both industry developers and open-source developers partaking in the experiment.}  
Therefore, we expect our results to generalize to both open and closed-source developers, \draft{5}{especially because the respondent's demographics are similar to those of the respondents to the Stack Overflow survey.} 
However, the most important threat to external validity we identify is related to the choice of vignettes. 
In the experiments, we only showed participants four instances, or vignettes, of technical debt, \draft{5}{and those instances all belonged one type of SATD, \emph{Poor implementation choices}. Therefore, the results of this experiment might not generalize over other types or categories of SATD, such as functional issues, or SATD described in issue tracking systems.}
However, we tried to minimize the threat by using four different vignettes instead of one, and therefore, trying to balance participant fatigue or disengagement with generalizability.

\paragraph{Construct Validity:} 
Most importantly, we recognize that \emph{Importance}, \emph{Urgency}, and \emph{Effort} as Likert scale questions are not operationalizations of priority used in practice by developers to score technical debt. However, we opted for these three constructs because they are easy to explain to participants and because there is no universally accepted construct to measure the priority of technical debt~\cite{Middleton:2018,Gavidia-Calderon:2021,Bellotti:2004}. 
\draft{5}{While the usage of these constructs might influence results, they allow us to measure perceived priority consistently. }

%% file: 2-6-related-work.tex
\section{Related Work}

The below section discusses the related work on SATD, its management, and sentiment analysis in software engineering.

\subsection{Self-Admitted Technical Debt}

Many different aspects of technical debt have been studied previously~\citep{Tom:2013}, such as, for instance, its management~\citep{Li:2015}, or the financial aspects~\cite{Ampatzoglou:2015}.
This work focuses on the self-admission of Technical Debt (SATD), often present in software systems as source-code comments~\citep{Maldonado:2015, Potdar:2014}.
As for technical debt, many different aspects of SATD have been studied. Such as the annotation practices of industry developers~\cite{Zampetti:2021}, the places where SATD has been described~\cite{Xavier:2020,Xavier:2022,Kashiwa:2022}, and the curation of datasets of SATD instances~\cite{Sridharan:2023,Guo:2021}.

Because of the negative connotations of technical debt, research has focused on the classification of SATD in source code. 
\citet{Liu:2018} found that a SATD detector based on text-mining techniques can automatically identify SATD. 
Meanwhile, \citet{Ren:2019} have used deep-learning models to classify whether source code snippets contain SATD, finding that these models, trained on large datasets, can be competitive. 
However, what automatic detection technique has the best predictive performance is not entirely known, as \citet{Guo:2021} found that simple rule-based detection techniques can still outperform existing deep-learning tools. 
\draft{5}{\citet{Cassee:2022} has found that different types of issues are described in SATD, including \emph{functional issues}, \emph{poor implementation choices}, and \emph{documentation issues}. Furthermore, from surveying developers, Cassee \etal find that a group of developers state they use negativity to prioritize SATD.}

Another often studied aspect of SATD is its removal; both \citet{Palomba:2017} and \citet{Peruma:2022} studied the link between refactorings and SATD.
\citet{Palomba:2017} found that code refactorings occur in locations where SATD is present, while \citet{Peruma:2022} finds that code refactoring 
often coincides with removing SATD. 
By studying a dataset of SATD instances \citet{Maldonado:2017}
found that 75\% of SATD instances are removed in subsequent source code revisions. 
Furthermore, \citet{Zampetti:2018} found that the removal of SATD is acknowledged in commit messages. 
Additionally, Zampetti \etal 
find that in 20\% to 50\% of cases, SATD is removed when entire methods or 
classes are removed. 
By surveying developers \citet{Tan:2021} found that self-removal of SATD is often a conscious decision. 
Moreover, they found that more experienced developers are more often concerned when it comes to the removal of SATD. 
However, none of the studies on the removal of SATD focus on whether the presence of negativity influences whether SATD is removed.

\subsection{Sentiment in Software Engineering}

Because this study focuses on the effect of sentiment on the prioritization of SATD, we describe literature related to the role of sentiment in software engineering. 
In general, sentiment analysis has been used to study a wide variety of software engineering activities~\cite{Lin:2022}, such as live meeting analysis~\citep{Hermann:2021} or to design and build recommender systems~\citep{Lin:2018}. 

Different studies have tried to understand whether expressions of negative sentiment are correlated with suboptimal development practices, such as unresolved issues, design smells, or bugs.
\citet{Valdez:2020} found that unresolved issues in Jira tend to express more negative sentiment than closed issues. 
Similarly, based on a study of issue reopenings \citet{Cheruvelil:2019} finds that issues that have been reopened once or more than once tend to have more comments expressing negative emotions. 
For commit messages \citet{Huq:2020} reports that commits related to bug fixing express more negative sentiment.
While for code reviews \citet{Asri:2019} finds that code reviews in which negative sentiment is expressed tend to take longer to complete. 
\citet{Olsson:2021} find that some design smells cause developers to feel negative emotions.

Wrt. positive sentiment, or the expression of positive emotions, \citet{Ortu:2015} has found that more positive sentiment in the description appears to correlate with a shorter issue resolution time. 
Similarly, \citet{Sanei:2021} has studied the sentiment in issue comments, finding that more positive comments correlate with faster resolution. 

After studying questions on Stack Overflow \citet{Calefato:2018} reports that successful questions on SO tend to be neutral, \ie express no positive or negative sentiment. Implying that neutral or more factual questions are more likely to receive a quicker response.

The existing studies focus on the impact of sentiment on software engineering through correlational analysis. 
However, the studies don't focus on causation (\ie ``did the expression of sentiment polarity cause the observed effect?''). 
Through the controlled experiment reported in this study, we further understand how negativity influences the perception of priority and explain how expressions of negativity impact software development.

%% file: 2-7-conclusion.tex
\section{Conclusion}

In this paper, we report on a controlled experiment to study whether developers interpret negativity in self-admitted technical debt as a proxy for priority. 
We exposed participants to instances of self-admitted technical debt with or without negativity and asked them to estimate the priority of each instance.
By analyzing the prioritization scores of the experiments using Bayesian statistics, we describe how negativity influences the perception of priority. 
Furthermore, to better describe the role negativity plays in the management of self-admitted technical debt, we asked respondents a series of questions about their perceptions towards the usage of negativity in SATD.

Based on the participation of 59 experienced industrial software engineers, we find that one-third to half of the developers are more likely to increase their estimation of priority if negativity is used to describe the SATD. \todo{conditional on own perception}
Specifically, they're between 1.5 and 2 times as likely to increase, as opposed to decrease, their prioritization score if negativity is present.
Secondly, we confirm previous findings and conclude that two-thirds of developers believe writing SATD in which negativity is expressed or interpreting negativity as a proxy for priority is unacceptable.
\draft{5}{Our findings show that developers who believe the usage of negativity as a proxy for priority is unacceptable still admit to writing negative SATD and use negativity to estimate priority. }

Based on the experiment, we learn how developers use negativity in SATD to communicate priority, in particular, we find that developers use negativity as an additional dimension, in addition to descriptions of technical issues. 
Furthermore, our results show why developers' expression of negativity is unavoidable and help explain the purpose of negative expressions. 
However, our results also show why using negativity to describe technical debt might not be advisable because of both developers' perceptions, and not all developers are influenced by it.